\documentclass[a4paper]{jpconf}
\usepackage{graphicx}
\begin{document}
\title{Nonequilibrium meson production in strong fields}

\author{L. Juchnowski$^a$, D. Blaschke$^{a,b,c}$, T. Fischer$^a$,  S.A. Smolyansky$^d$}

\address{$^a$ Institute of Theoretical Physics, University of Wroclaw, 50-204 Wroclaw, Poland,\\
$^b$ Bogoliubov Laboratory for Theoretical Physics, JINR Dubna, 141980 Dubna, Russia\\
$^c$ National Research Nuclear University (MEPhI), 
115409 Moscow, Russia\\
$^d$ Department of Physics, Saratov State University, 410026 Saratov, Russia}

\ead{lukasz.juchnowski@ift.uni.wroc.pl}

\begin{abstract}
We develop a kinetic equation approach to nonequilibrium pion and sigma meson production in a time-dependent, chiral symmetry breaking field (inertial mechanism).
We investigate the question to what extent the low-momentum pion enhancement observed in heavy-ion collisions at CERN - LHC 
can be addressed within this formalism.
In a first step, we consider the inertial mechanism for nonequilibrium production of $\sigma-$mesons and their simultaneous decay into pion pairs for two cases of $\sigma$ mass evolution. 
The resulting pion distribution shows a strong low-momentum enhancement which can be approximated by a thermal Bose distribution with a chemical potential that appears as a trace of the nonequilibrium process of its production. 
\end{abstract}

\section{Kinetic theory for inertial particle production}
The formulation of a kinetic theory approach to the problem of nonequilibrium particle production in strong fields has been advanced recently in studies of the dynamical Schwinger effect for $e^{+}e^{-}$ plasma creation in high-intensity lasers  \cite{Blaschke:2008wf,Blaschke:2014fca}. 
Strong and time-dependent fields govern also the particle production in ultrarelativistic heavy-ion collisions.
Here the kinetic theory approach has been developed, e.g., for studying the role of time-dependent masses (the so-called "inertial mechanism" \cite{Filatov:2008}) in the course of the chiral symmetry breaking transition for pion production \cite{Filatov:2008zz} and photon production \cite{Michler:2012mg}.
In the present work we develop the kinetic approach to nonequilibrium pion production in a time-dependent chiral symmetry breaking homogeneous field and investigate the question to what extent the low-momentum pion enhancement observed in heavy-ion collisions at CERN - LHC being discussed as Bose-Einstein condensation of pions \cite{Begun:2015ifa} can be described within this formalism.
To this end we set up a detailed study of the three main processes that are intertwined in this case:
(a) the nonequilibrium $\sigma-$meson production in the time-dependent external field, 
(b) the $\sigma\to \pi\pi$ decay and (c) the $\pi\pi$ rescattering and formation of the Bose condensate
\cite{Semikoz:1994zp,Voskresensky:1996ur}.
The results shall be compared with the effect observed at the LHC. Here we report on steps (a) and (b).

In order to address the question of the $\sigma$ and $\pi$ meson production in a heavy-ion collision we consider the simplified situation of one species of pions only, i.e. we neglect the isospin degree of freedom. We describe the evolution of the single-particle distribution functions $f_\sigma(t,\vec{x},\vec{p})$ and $f_\pi(t,\vec{x},\vec{p})$ 
as solutions of  the coupled Boltzmann equations for these relativistic bosons with the dispersion relations,
\begin{equation}
\label{dispersion}
\omega_\sigma(t,\vec{p}) = \sqrt{m_\sigma(T(t))^2 + \vec{p}^2}\;, \\
\quad \omega_\pi(\vec{p}) = \sqrt{m_\pi^2 + \vec{p}^2}\;, \quad  m_\pi = 140 \mbox{ MeV}~.
\end{equation}
For the evolution of the mass of the $\sigma$, we apply the following expression
\begin{equation}
\label{eq:sigmamass}
 m_\sigma(T(t)) = [m_\sigma(0)-m_\pi] \sqrt{1-\frac{T(t)}{T_c}}+m_\pi\;, \quad T(t) = \frac{T_0~t_0}{t}\;, \quad t\ge t_0\;,
\end{equation}
where $ T_c = 170$ MeV is critical temperature for the chiral transition and $m_\sigma(0)$ is the vacuum 
$\sigma$ mass. 
The time at the begin of the 
3-dimensional spherical expansion is $t_0=9$ fm/c, corresponding to a Hubble flow velocity of 
$v_R=0.72$ c for gold nuclei with radius $R_0=6.5$ fm. 
The initial temperature $T_0=T(t_0)$ is taken to be $T_0=T_c$.

Here we assume spatial homogeneity, i.e. the distribution functions have no dependence on position and despite their momentum dependence we neglect derivative terms so $df/dt = \partial f/\partial t$. In our simplified model the evolution of $\sigma$ and $\pi$ are dominated by 
$\sigma$ production in the evolving chiral condensate (inertial mechanism) and the subsequent decay $\sigma \to \pi\pi$.  
The Boltzmann transport equation for $\sigma$ where the rescattering of pions is not yet considered at this step, reads
\begin{eqnarray}
&&\frac{\partial f_\sigma}{\partial t}(t,\vec{p}_\sigma)
\nonumber
=
\frac{\Delta_\sigma(t,\vec{p} _\sigma)}{2}\int_{t_0}^t dt' \Delta_\sigma(t',\vec{p} _\sigma) \left(1+f_\sigma(t',\vec{x},\vec{p} _\sigma)\right)
\cos\left(2\theta_\sigma(t,t',\vec{p} _\sigma)\right)
\nonumber
\\
&&+
\left(1+f_\sigma(t,\vec{p} _\sigma)\right)
\left(
\int \frac{d^3p_1}{(2\pi)^3 2\omega_1}\frac{d^3p_2}{(2\pi)^3 2\omega_2} \Gamma_{\pi\pi\rightarrow\sigma}(\vec{p} _\sigma,\vec{p}_1,\vec{p}_2)
f_\pi(t,\vec{p}_1) f_\pi(t,\vec{p}_2)
\right)
\nonumber
\\
&&- 
f_\sigma(t,\vec{p}_\sigma)
\left(
\int \frac{d^3p_1}{(2\pi)^3 2\omega_1}\frac{d^3p_2}{(2\pi)^3 2\omega_2} \Gamma_{\sigma\rightarrow\pi\pi}(\vec{p} _\sigma,\vec{p}_1,\vec{p}_2)
\left(1+f_\pi(t,\vec{p}_1)\right)\left(1+f_\pi(t,\vec{p}_2)\right)
\right)
\;.
\label{eq:sigma_transport}
\end{eqnarray}
Note that the source term for $\sigma$ production occurs just due to the time dependence of the $\sigma$ dispersion law (\ref{dispersion}) and works in the absence of pions. 
It is the first term at the right hand side of Eq.~\ref{eq:sigma_transport}, with the following definitions

\begin{equation}
\label{eq:Delta}
\Delta_\sigma(t,\vec{p} _\sigma)=\frac{m_\sigma}{\omega_\sigma^2}\frac{\partial m_\sigma}{\partial t}\;, \quad \theta_\sigma(t,t',\vec{p} _\sigma) = \int_{t'}^t dt'' \omega_\sigma(t'',\vec{p} _\sigma)\;.
\end{equation}

The last two terms in Eq.~\ref{eq:sigma_transport} are due to the $\sigma\to \pi\pi$ decay and regeneration $\pi\pi\to\sigma$. 
For the pions, the dispersion law is time-independent for $t>t_0$ so that there is no inertial production mechanism
\begin{eqnarray}
&&\frac{\partial f_\pi}{\partial t}(t,\vec{p}_1)
=\nonumber
\\
&&=
\left(1+f_\pi(t,\vec{p_1})\right)
\left(
\int \frac{d^3p _\sigma}{(2\pi)^3 2\omega_\sigma}\frac{d^3p_2}{(2\pi)^3 2\omega_2} \Gamma_{\sigma\rightarrow\pi\pi}(\vec{p} _\sigma,\vec{p_1},\vec{p_2})
\left(1+f_\pi(t,\vec{p_2})\right)
f_\sigma(t,\vec{p}_\sigma)
\right)
\nonumber
\\
&&-
f_\pi(t,\vec{p_1})
\left(
\int \frac{d^3p_\sigma}{(2\pi)^3 2\omega_\sigma}\frac{d^3p_2}{(2\pi)^3 2\omega_2} 
\Gamma_{\pi\pi\rightarrow\sigma}(\vec{p} _\sigma,\vec{p_1},\vec{p_2})
f_\pi(t,\vec{p_2})
\left(1+f_\sigma(t,\vec{p} _\sigma)\right)
\right)
\;.
\label{eq:pi_transport}
\end{eqnarray}
\newline
%
For the $\sigma$ decay and regeneration we assume a constant matrix element $|M|^2={\rm const}$ so that the momentum dependence of 
$\Gamma_{\sigma\to\pi\pi}$ is simply given by the momentum conserving delta-function 
%

\begin{eqnarray}
\Gamma_{\sigma\rightarrow\pi\pi}(\vec{p}_\sigma,\vec{p}_1,\vec{p}_2)
=(2\pi)^4 \delta^4(p_\sigma-p_1-p_2) \vert M \vert^2
\rightarrow
(2\pi)^4  \delta(w_\sigma-w_1-w_2)\delta^3(\vec{p}_\sigma-\vec{p}_1-\vec{p}_2)
\;.
\label{eq:Gamma_delta}
\end{eqnarray}

\begin{figure}[!htb]
\includegraphics[width=\textwidth]{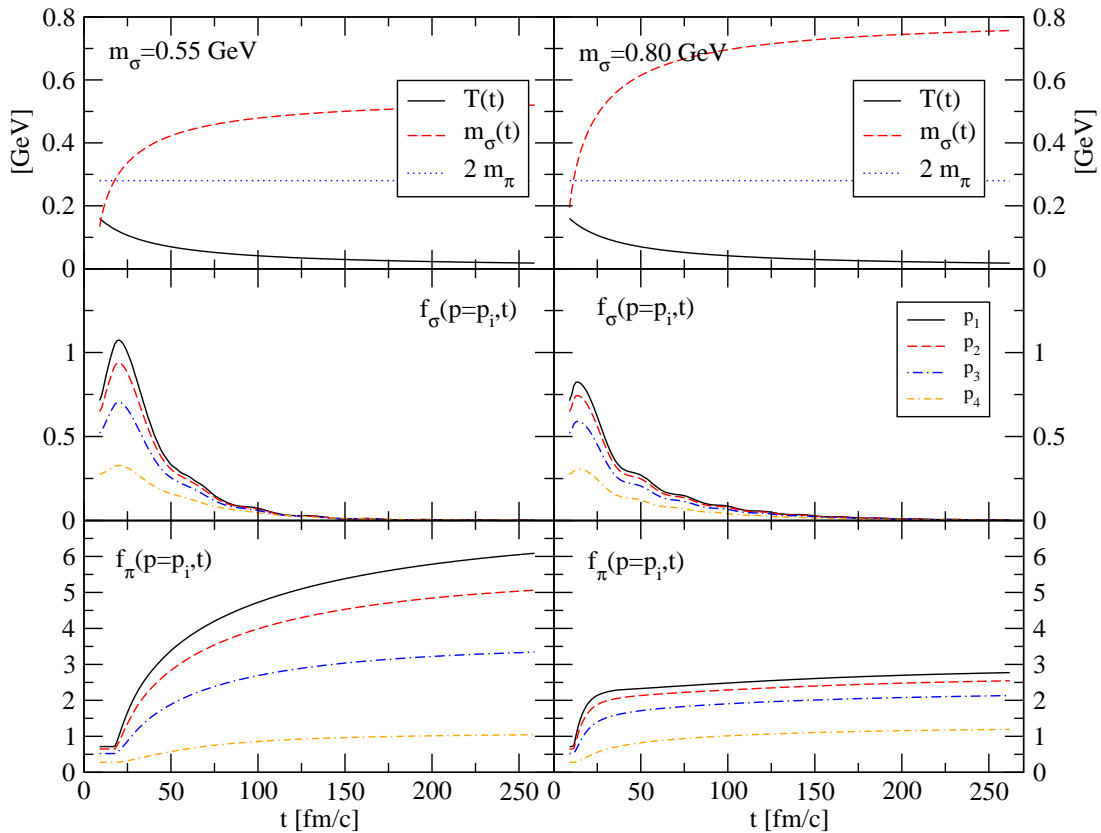}
\caption{\label{fig:evolution}Coupled $\sigma-\pi$ kinetics in an evolving scalar background field (upper panels, red dashed lines) with a vacuum $\sigma$ mass term of $550$ MeV (left panels) and $800$ MeV (right panels).
Evolution of the $\sigma$ (middle panels) and $\pi$ (bottom panels) distribution function at four selected momenta $p_1 = 5 $ MeV, $p_2 = 50 $ MeV, $p_3 = 100 $ MeV, $p_4 = 200 $ MeV. 
Pion production is active only above the sigma-2pi threshold when  $m_\sigma>2~m_\pi$.}
\end{figure}

\section{Results}
In order to apply the inertial mechanism of meson production described by the coupled kinetic equations (\ref{eq:sigma_transport})  and (\ref{eq:pi_transport}) to the case of heavy-ion collisions at LHC, we consider a scenario 
where the initial state is described by thermal equilibrium distributions of $\pi$ and $\sigma$ mesons with degenerate mass $m_\pi(T_0)=m_\sigma(T_0)=140$ MeV (chiral symmetry) at $T_0=T_c=170$ MeV 
\begin{eqnarray}
f_{i}(t_0,\vec{p})=g_{i} \left[\exp(\sqrt{p^2+m_i^2(T_0)}/T_0) -1\right]^{-1}~,~~i=\pi,\sigma~.
\end{eqnarray}
In the subsequent evolution the $\sigma$ mass departs from the pion mass and rises towards its vacuum value $m_\sigma(0)$ (chiral symmetry breaking) while the
the pion mass keeps the value at the onset of the chiral symmetry breaking due to chiral protection. 

According to our model the chiral transition takes place because of  very  fast 3-dimensional expansion (and thus dilution and cooling) of the fireball. 
Therefore it is reasonable to assume that the process $\pi + \pi \rightarrow \sigma$ is strongly suppressed. 
In such a case  one can disregard the terms containing $\Gamma_{\pi\pi\to\sigma}$ in Eqs.~(\ref{eq:sigma_transport}) and (\ref{eq:pi_transport}). 
The resulting system of kinetic equations to be solved is given by

\begin{eqnarray}
\frac{\partial f_\sigma}{\partial t}(t,p_\sigma)
&=&
\frac{\Delta_\sigma(t,p _\sigma)}{2}\int_{t_0}^t dt' \Delta_\sigma(t',p_\sigma) \left(1+f_\sigma(t',p _\sigma)\right)
\cos\left(2\theta_\sigma(t,t',p _\sigma)\right)
\nonumber
\\
&-&
\frac{1}{8\pi}
f_\sigma(t,p_\sigma)
\frac{1}{p_\sigma w_\sigma}
\int_{p_1^-}^{p_1^+} p_1 dp_1 
\frac{\vert M \vert^2}{w_1}
\left(1+f_\pi(t,p_1)\right)\left(1+f_\pi(t,p_2(z_0,p_1;p_\sigma)\right)\,,
\label{eq:fullsigma}
\end{eqnarray}

\begin{eqnarray}
\frac{\partial f_\pi}{\partial t}(t,p_1)
&=&
\frac{1}{8\pi}
\left(1+f_\pi(t,p_1)\right)
\frac{1}{p_1 w_1}
\int_{p_\sigma^-}^{p_\sigma^+} p_\sigma dp_\sigma 
\frac{\vert M \vert^2}{w_\sigma}
f_\sigma(t,p_\sigma)\left(1+f_\pi(t,p_2(z_0,p_\sigma;p_1))\right)\,,
\label{eq:pi_transport_red}
\end{eqnarray}
where 
\begin{eqnarray}
&&
p_1^\pm = \frac{1}{2}\left\vert p_\sigma\pm w_\sigma\sqrt{1-\frac{4 m_\pi^2}{m_\sigma^2}}\right\vert ,\quad\quad p_\sigma^\pm = \frac{m_\sigma^2}{m_\pi^2}\frac{1}{2}
\left\vert p_1 \pm w_1 \sqrt{1-\frac{4 m_\pi^2}{m_\sigma^2}} \right\vert
\;.
\end{eqnarray}

The results for the evolution of the distribution functions due to the coupled $\sigma-\pi$ kinetics in the evolving scalar background field are presented in Fig.~\ref{fig:evolution}. 
One can notice that the $\sigma$ distributions in the middle panels of Fig.~\ref{fig:evolution} show oscillatory behaviour which is common for the kinetic approach to particle production and has been discussed also in the context of the  dynamical Schwinger effect in lasers 
\cite{Blaschke:2008wf,Blaschke:2014fca}. 

\begin{figure}[!th]
\begin{minipage}[l]{0.6\textwidth}
\includegraphics[width=\textwidth]{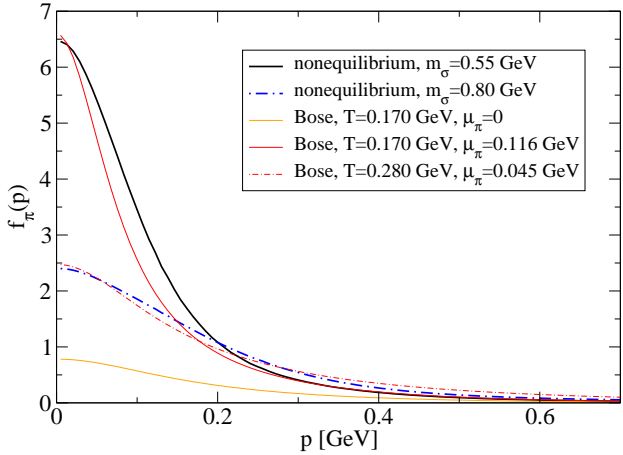}
\end{minipage}\hfill
\begin{minipage}[r]{0.4\textwidth}
\caption{\label{fig:thermal}
Nonequilibrium distribution functions for pions for two different cases of vacuum $\sigma$ mass: $550$ MeV (black solid lines) and $800$ MeV (blue dash-dotted lines) together with their approximations by thermal Bose distributions with ($T=170$ MeV, $\mu_\pi=116$ MeV) and ($T=280$ MeV, $\mu_\pi=45$ MeV), resp. The initial thermal pion distribution with temperature $T=170$ MeV is shown by the orange solid line.}
\end{minipage}
\end{figure}

Results presented in Fig.~\ref{fig:thermal} show $f_\pi(t_f,\vec{p})$ at sufficiently late times $t_f>250$ fm/c plotted together with the Bose distribution of the initial state. 
In this stage $f_\sigma(t_f,\vec{p})$ is already negligible because all $\sigma$ mesons have decayed.
In the pion channel we obtain a strong enhancement at low momenta which stems from the decay of the 
$\sigma$ mesons produced by the 
inertial mechanism. 
In order to quantify this effect we introduce a pion enhancement ratio 
\begin{equation}
r_\pi = \int_0^\infty dp p^2 f_\pi(t_f,\vec{p}) / \int_0^\infty dp p^2 f_\pi(t_0,\vec{p})
\end{equation}
and obtain for it the value $r_\pi=2.98$ ($r_\pi=3.18$) in the scenario with $m_{\sigma}(0)=0.55$ GeV
($m_{\sigma}(0)=0.80$ GeV). 
This pattern of low-momentum pion enhancement and its approximate description by a pion chemical potential has already been noticed for early CERN SPS experiments \cite{Kataja:1990tp}.
It was subsequently disregarded in favour of a description by resonance decays which, however, 
appear to be insufficient for explaining the magnitude of the effect as observed recently by the ALICE experiment at LHC. 
The effect of pion enhancement from the inertial mechanism as discussed above has the right order of magnitude to explain the observation.
It thus qualifies as a possible microscopic explanation for the nonequilibrium origin.
Very roughly it can be captured by assigning a nonequilibrium chemical potential to the pions.
Thermalization should be accomplished by including elastic $\pi-\pi$ rescattering processes.


\section{Conclusion and outlook}
In the present work we have developed further the kinetic approach to nonequilibrium pion production 
by the inertial mechanism in a time-dependent chiral symmetry breaking homogeneous field. 
We have addressed the question to what extent the low-momentum pion enhancement observed in 
heavy-ion collisions at CERN - LHC being discussed as Bose-Einstein condensation of pions [5] can be described within this formalism. For simplicity we have neglected the isospin of pions here.
Along the lines of this project, we have performed the first two steps (a) and (b) of a detailed study of the three main processes that are intertwined in this case:
(a) the nonequilibrium $\sigma-$meson production in the time-dependent external field, 
(b) the $\sigma\to \pi\pi$ decay. 
The step (c) consisting in the inclusion of $\pi\pi$ rescattering and formation of the Bose condensate
\cite{Semikoz:1994zp,Voskresensky:1996ur} as well as the comparison of the obtained results with the effect observed at the LHC is subject to current research.
At the present stage we can conclude that the distribution function $f_\pi(t)$ depends strongly  on magnitude, shape and duration of the chiral symmetry breaking (inertial) source term $\Delta_\sigma(t)$. 
The distribution function $f_\pi(t)$ before $\pi-\pi$ rescattering can roughly be approximated 
(for the more realistic case $m_{\sigma}(0)=0.55$ GeV) by a Bose distribution with a nonequilibrium pion chemical potential $\mu_\pi=116$ MeV and a freeze-out temperature $T=170$ MeV.

\ack
We are grateful to V. Begun, W. Florkowski, P.M. Lo, G. R\"opke, L. Turko and D.N. Voskresensky for their enlightening discussions. We thank also E.-M. Ilgenfritz and A. Tawfik for their continued interest in the progress of this project.
This research is supported in part by the Polish Narodowe Centrum Nauki (NCN) under grant number UMO-2014/15/B/ST2/03752 (L.J. and D.B.) and UMO-2013/11/D/ST2/02645 (T.F.).

\end{document}